\documentclass[journal,10pt, twocolumn]{IEEEtran}
\usepackage{amsfonts}
\usepackage{amsfonts}
\usepackage{amsfonts}
\usepackage{amsfonts}
\usepackage{amsfonts}
\usepackage[latin1]{inputenc}
\usepackage[usenames]{color}
\usepackage[dvips]{pstcol}
\usepackage{amssymb}
\usepackage{latexsym}
\usepackage{mathrsfs}
\usepackage{amsfonts}
\usepackage{amsbsy}
\usepackage{amsmath}
\usepackage{graphicx}
\usepackage{times}
\usepackage{epsfig,epsf,times,amssymb,amsmath}

\begin{document}
\title{\huge Low-Complexity Reduced-Rank Beamforming Algorithms}

\author{Lei~Wang and
        ~Rodrigo~C.~de~Lamare
\thanks{The authors are with the Department
of Electronics, The University of York, York, YO10 5DD, U.K.
(e-mail:
lw517@york.ac.uk, rcdl500@ohm.york.ac.uk).}\\
\thanks{This work is supported by the Engineering and Physical Sciences Research Council (EPSRC), U.K., EPSRC Reference EP/H011544/1.}}

\maketitle

\begin{abstract}
A reduced-rank framework with set-membership filtering (SMF)
techniques is presented for adaptive beamforming problems
encountered in radar systems. We develop and analyze stochastic
gradient (SG) and recursive least squares (RLS)-type adaptive
algorithms, which achieve an enhanced convergence and tracking
performance with low computational cost as compared to existing
techniques. Simulations show that the proposed algorithms have a
superior performance to prior methods, while the complexity is
lower.
\end{abstract}


\section{Introduction}

With the development of array signal processing techniques,
beamforming has long been investigated for numerous applications in
radar, sonar, seismology, and wireless communications
\cite{Johnson}, \cite{Trees}. The most well-known beamforming
technique is the optimal linearly constrained minimum variance
(LCMV) beamformer \cite{Frost,delamaretsp}. It exploits the
second-order statistics of the received vector to minimize the array
output power while constraining the array response in the direction
of the signal of interest (SOI) to be constant. In general, the
constraint corresponds to prior knowledge of the direction of
arrival (DOA) of the SOI.

Many adaptive algorithms have been reported for the implementation
of the LCMV beamformer, ranging from the low-complexity stochastic
gradient (SG) algorithm to the more complex recursive least squares
(RLS) algorithm \cite{Haykin}. According to the parameter estimation
strategy of the algorithm, the SG and RLS algorithms can be included
in the class of full-rank processing techniques\cite{Honig2}. The
full-rank adaptive algorithms usually require a large number of
snapshots to reach the steady-state when the number of elements in
the beamformer is large, and the resulting convergence speed reduces
significantly. In dynamic scenarios (e.g., when interferers enter or
exit a given system), filters with many elements show a poor
tracking performance when dealing with signals embedded in
interference and noise. These situations are quite relevant in
defence systems such as radar. {  Other strategies for interference
suppression coming from the antenna community include the recent
work by Massa {\it et al.} \cite{massa} that introduces a dynamic
thinning strategy, the work by D'Urso {\it et al.} \cite{durso} that
considers a hybrid optimization procedure that adjusts both
clustering into subarrays and excitations of the subarrays, the
contribution of Haupt \cite{haupt1} which uses subarrays a hybrid
genetic algorithm to optimize the size of the subarrays their
weights, the method of Haupt and Aten \cite{haupt2} which employs a
genetic algorithm to optimize the orientation of each dipole in an
array, and the technique by Haupt {\it et al.} \cite{haupt3} that
uses partial adaptation of the beamforming weights.}

These problems motivate us to investigate a more effective signal
processing approach known as reduced-rank signal processing, which
allows a designer to address the drawbacks of full-rank algorithms.
The idea is to employ a transformation matrix that projects the
received signal onto a lower dimensional subspace, and then the
reduced-rank filter optimization occurs within this subspace. This
has the advantage of improving the convergence and tracking
performance. The advantage is more obvious when the number of sensor
elements in the array is large. Well-known reduced-rank schemes
include the multistage Wiener filter (MSWF)
\cite{Goldstein}-\cite{Lamare}, the auxiliary vector filtering (AVF)
\cite{Pados},\cite{Mathews} the joint iterative optimization (JIO)
\cite{jio_spl}-\cite{jio_stap} and the joint interpolation,
decimation and filtering (JIDF)-based approaches
\cite{jidf}-\cite{barc}. { They employ different procedures to
construct the transformation matrix and to estimate the parameters.
A common problem of these reduced-rank schemes is the relatively
high computational load required to
compute the transformation matrix.} 

An efficient approach to reducing the computational complexity is to
employ a set-membership filtering (SMF) technique
\cite{Gollamudi,Nagaraj} for the beamformer design. The SMF
specifies a predetermined bound on the magnitude of the estimation
error or the array output and performs data-selective updates to
estimate the parameters. It involves two steps: $1)$ information
evaluation (depending on the predetermined bound) and $2)$ parameter
update (depending on step $1)$). If the parameter update does not
occur frequently, and the information evaluation does not require
much complexity, the overall complexity can be substantially
reduced. The well-known SMF algorithms include the SG-based
algorithms in \cite{Gollamudi} and the RLS-based algorithms in
\cite{Nagaraj}, \cite{Guo}. These algorithms are examples of the
application of the SMF technique in the full-rank signal processing
context.

The objective of this paper is to introduce a constrained
reduced-rank framework and algorithms for achieving a superior
convergence and tracking performance with significantly lower
computational cost comparable with their reduced-rank counterparts.
We consider reduced-rank LCMV designs using the SMF concept that
imposes a bounded constraint on the array output and the JIO
strategy. The joint optimization of the transformation matrix and
the reduced-rank filter are then performed for beamforming. The
reduced-rank parameters only update if the bounded constraint cannot
be satisfied. This partial update plays a positive role in
increasing the convergence speed. The updated parameters belong to a
set of feasible solutions. Considering the fact that the
predetermined bound degrades the performance of the SMF technique
due to the lack of knowledge of the environment, we utilize a
parameter-dependent time-varying bound instead to guarantee a good
performance. Related work can be found in \cite{Guo2},
\cite{Lamare4} but only focuses on the full-rank signal processing
context. In this paper, we introduce this technique into the
reduced-rank signal processing context. The proposed framework,
referred here as JIO-SM, inherits the positive features of the
reduced-rank JIO schemes that jointly and iteratively exchange
information between the transformation matrix and the reduced-rank
filter, and performs beamforming using the SMF data-selective
updates. We propose constrained reduced-rank SG-based and RLS-based
adaptive algorithms, namely, JIO-SM-SG and JIO-SM-RLS, for the
design of the proposed beamformer. A discussion on the properties of
the developed algorithms is provided. Specifically, a complexity
comparison is presented to show the advantages of the proposed
algorithms over their existing counterparts. A mean-squared error
(MSE) expression to predict the performance of the proposed
JIO-SM-SG algorithm is derived. We also analyze the properties of
the optimization problem by employing the SMF constraint.
Simulations are provided to show the performance of the proposed and
existing algorithms.

The remainder of this paper is organized as follows: we outline a
system model for beamforming in Section II. Based on this model, the
full-rank and the reduced-rank LCMV beamformer are reviewed. The
novel reduced-rank framework based on the JIO scheme and the SMF
technique is presented in Section III, and the proposed adaptive
algorithms are detailed in Section IV. A complexity study and the
related analyses of the proposed algorithms are carried out in
Section V. Simulation results are provided and discussed in Section
VI, and conclusions are drawn in Section VII.

\section{System Model and LCMV Beamformer Design}
In this section, we describe a system model to express the array
received vector. Based on this model, the full-rank and the
reduced-rank LCMV beamformers are introduced.

\subsection{System Model}
Let us suppose that $q$ narrowband signals impinge on a uniform
linear array (ULA) of $m$ ($m\geq q$) sensor elements. The sources
are assumed to be in the far field with DOAs
$\theta_{0}$,\ldots,$\theta_{q-1}$. The received vector $\boldsymbol
x\in\mathbb C^{m\times 1}$ can be modeled as
\begin{equation} \label{1}
\centering {\boldsymbol x}={\boldsymbol {A}}({\boldsymbol
{\theta}}){\boldsymbol s}+{\boldsymbol n},
\end{equation}
where
$\boldsymbol{\theta}=[\theta_{0},\ldots,\theta_{q-1}]^{T}\in{\mathbb{R}}^{q
\times 1}$ is the vector with the signals' DOAs, ${\boldsymbol
A}({\boldsymbol {\theta}})=[{\boldsymbol
a}(\theta_{0}),\ldots,{\boldsymbol a}(\theta_{q-1})]\in\mathbb{C}^{m
\times q}$ comprises the normalized signal steering vectors
${\boldsymbol a}(\theta_{k})=[1, e^{-2\pi
j\frac{u}{\lambda_{\textrm{c}}}cos{\theta_{k}}},\ldots$, $e^{-2\pi
j(m-1)\frac{u}{\lambda_{\textrm{c}}}cos{\theta_{k}}}]^{T}\in\mathbb{C}^{m
\times 1}$, $(k=0,\ldots,q-1)$, where $\lambda_{\textrm{c}}$ is the
wavelength and $u$ ($u=\lambda_{\textrm{c}}/2$ in general) is the
inter-element distance of the ULA. To avoid mathematical
ambiguities, the steering vectors $\boldsymbol a(\theta_{k})$ are
assumed to be linearly independent, ${\boldsymbol s}\in
\mathbb{C}^{q\times 1}$ is the source data vector, ${\boldsymbol
n}\in\mathbb{C}^{m\times 1}$ is the noise vector, which
is assumed to be a zero-mean spatially and Gaussian process, and
$(\cdot)^{T}$\ stands
for transpose. 

\subsection{Full-rank LCMV Beamformer Design}
The full-rank LCMV beamformer design is equivalent to determining a
set of filter parameters $\boldsymbol w=[w_1, \ldots,
w_m]^T\in\mathbb C^{m\times1}$ that provide the array output
$y=\boldsymbol w^H\boldsymbol x$, where $(\cdot)^H$ represents
Hermitian transpose. The filter parameters are calculated by solving
the following optimization problem:
\begin{equation}\label{2}
\textrm{minimize}~~\mathbb E[|y|^2]=\mathbb E[|\boldsymbol
w^H\boldsymbol x|^2],~~~~~\textrm{subject~to}~~\boldsymbol
w^H\boldsymbol a(\theta_0)=\gamma,
\end{equation}
where $\boldsymbol a(\theta_0)$ is the full-rank steering vector of
the SOI and $\gamma$ is a constant. The objective of (\ref{2}) is to
minimize the array output power while maintaining the contribution
from $\theta_0$ constant.

The solution of the LCMV optimization problem is
\begin{equation}\label{3}
\boldsymbol w_{\textrm{opt}}=\frac{\gamma\boldsymbol
R^{-1}\boldsymbol a(\theta_0)}{\boldsymbol a^H(\theta_0)\boldsymbol
R^{-1}\boldsymbol a(\theta_0)},
\end{equation}
where $\boldsymbol R=\mathbb E[\boldsymbol x\boldsymbol
x^H]\in\mathbb C^{m\times m}$ is the received data covariance
matrix. The filter $\boldsymbol w$ can be estimated in an adaptive
way via SG or RLS algorithms, where $\boldsymbol R$ is calculated by
its sample estimate. However, their convergence and tracking
performance depends on the filter length $m$, and degrades when $m$
is large \cite{Haykin}, \cite{jio_lcmv_esp}.

\subsection{Reduced-rank LCMV Beamformer Design}
An important feature of the reduced-rank schemes is to construct a
transformation matrix $\boldsymbol T_r\in\mathbb C^{m\times r}$ that
performs the dimensionality reduction that projects the full-rank
received vector onto a lower dimension, which is given by
\begin{equation}\label{4}
\bar{\boldsymbol x}=\boldsymbol T_r^H\boldsymbol x,
\end{equation}
where $\bar{\boldsymbol x}$ denotes the reduced-rank received vector
and $r(1\leq r\leq m)$ is the rank. In what follows, all
$r$-dimensional quantities are denoted with a ``bar".

The reduced-rank LCMV beamformer estimates the parameters
$\bar{\boldsymbol w}=[\bar{w}_1, \ldots, \bar{w}_r]^T\in\mathbb
C^{r\times1}$ to generate the array output $y=\bar{\boldsymbol
w}^H\bar{\boldsymbol x}$. The reduced-rank filter is designed by
solving the optimization problem:
\begin{equation}\label{5}
\begin{split}
\textrm{minimize}~~\mathbb E[|y|^2]& =\mathbb E[|\bar{\boldsymbol
w}^H\bar{\boldsymbol x}|^2],\\
\textrm{subject~to} ~~\bar{\boldsymbol w}^H\bar{\boldsymbol
a}(\theta_0)&=\gamma,
\end{split}
\end{equation}
where $\bar{\boldsymbol a}(\theta_0)=\boldsymbol T_r^H\boldsymbol
a(\theta_0)$ is the reduced-rank steering vector with respect to the
SOI. The solution of the reduced-rank LCMV optimization problem is
\begin{equation}\label{6}
\begin{split}
\bar{\boldsymbol w}_{\textrm{opt}} =\frac{\gamma\bar{\boldsymbol
R}^{-1}\bar{\boldsymbol a}(\theta_0)}{\bar{\boldsymbol
a}^H(\theta_0)\bar{\boldsymbol R}^{-1}\bar{\boldsymbol
a}(\theta_0)},
\end{split}
\end{equation}
where $\bar{\boldsymbol R}=\mathbb E[\bar{\boldsymbol
x}\bar{\boldsymbol x}^H]=\boldsymbol T_r^H\boldsymbol R\boldsymbol
T_r\in\mathbb C^{r\times r}$ is the reduced-rank data covariance
matrix. The MSWF \cite{Honig}, \cite{Lamare}, the AVF \cite{Pados},
and the JIO \cite{jio_spl} are effective reduced-rank schemes to
construct the transformation matrix aided by SG-based or RLS-based
adaptive algorithms for parameter estimation. However, there is a
number of problems and limitations with the existing techniques. The
computational complexity of algorithms dealing with a large number
of parameters can be substantial. It is difficult to predetermine
the step size or the forgetting factor values to achieve a
satisfactory tradeoff between fast convergence and misadjustment
\cite{Trees}. The RLS algorithm present problems with numerical
stability and divergence \cite{Haykin}. Furthermore, the
computational cost is high if the transformation matrix has to be
updated for each snapshot.

\section{Proposed JIO-SM Framework}

\begin{figure}[htb]
\begin{minipage}[h]{1.0\linewidth}
  \centering
  \centerline{\epsfig{figure=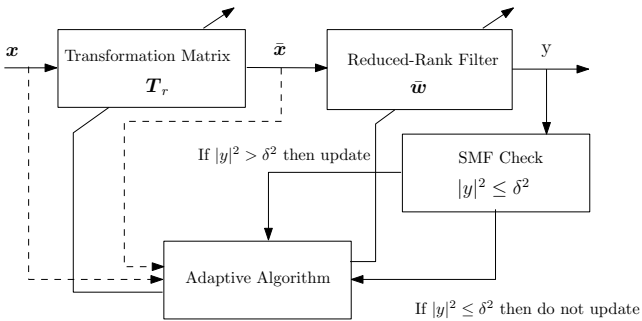,scale=0.625}}
  \vspace{-0.25em}\caption{Proposed reduced-rank JIO-SM structure.} \label{fig:jio_sm1}
\end{minipage}
\end{figure}

In order to address some of the problems stated in Section II, we
introduce a new constrained reduced-rank framework to address them
by combining the SMF techniques with the reduced-rank JIO scheme, as
depicted in Fig. \ref{fig:jio_sm1}. In this structure, the
transformation matrix is constructed using a bank of $r$ full-rank
filters $\boldsymbol t_j=[t_{1,j}, t_{2,j}, \ldots,
t_{m,j}]^T\in\mathbb C^{m\times1}$, ($j=1, \ldots, r$), as given by
$\boldsymbol T_r=[\boldsymbol t_1, \boldsymbol t_2, \ldots,
\boldsymbol t_r]$. The transformation matrix processes the received
vector $\boldsymbol x$ for reducing the dimension, and retains the
key information of the original signal in the generated reduced-rank
received vector $\bar{\boldsymbol x}$. The reduced-rank filter
$\bar{\boldsymbol w}$ then computes the output $y$.

For the JIO scheme, the reduced-rank adaptive algorithms
\cite{jio_spl} are developed to update $\boldsymbol T_r$ and
$\bar{\boldsymbol w}$ with respect to each time instant ``$i$". In
the proposed JIO-SM structure, the SMF check is embedded to specify
a time-varying bound $\delta$ (with respect to $i$) on the amplitude
of the array output $y$. The time-varying bound is related to the
previous transformation matrix and the reduced-rank weight vector.
The parameter update is only performed if the constraint on the
bound $|y|^2\leq|\delta|^2$ cannot be satisfied. At each time
instant, some valid pairs $\{\boldsymbol T_r, \bar{\boldsymbol w}\}$
are consistent with the bound. Therefore, the solution to the
proposed JIO-SM scheme is a set in the parameter space. Some pairs
of $\{\boldsymbol T_r, \bar{\boldsymbol w}\}$ even satisfy the
constrained condition with respect to different received vectors for
different ``$i$". Thus, the proposed scheme only takes the
data-selective updates and ensures all the updated pairs satisfy the
constraint for the current time instant. In comparison, the
conventional full-rank or reduced-rank filtering schemes only
provide a point estimate with respect to the received vector for
each time instant. This estimate may not satisfy the condition with
respect to other received vectors (at least before the algorithm
achieves the steady-state). Compared with the existing SMF
techniques \cite{Gollamudi}-\cite{Guo}, the proposed scheme takes
both $\boldsymbol T_r$ and $\bar{\boldsymbol w}$ into consideration
with respect to the bounded constraint in order to promote an
exchange of information between them. This procedure ensures the key
information of the original signal to be utilized more effectively.

Let $\mathcal {H}_i$ denote the set containing all the pairs of
$\{\boldsymbol T_r, \bar{\boldsymbol w}\}$ for which the associated
array output at time instant $i$ is upper bounded in magnitude by
$\delta$, which is
\begin{equation}\label{7}
\mathcal {H}_i=\big\{\boldsymbol T_r\in\mathbb C^{m\times r},
\bar{\boldsymbol w}\in\mathbb C^{r\times1}:~|y|^2\leq\delta^2\big\},
\end{equation}
where $\mathcal H_i$ is bounded by a set of hyperplanes that
correspond to the pairs of $\{\boldsymbol T_r, \bar{\boldsymbol
w}\}$. The set $\mathcal H_i$ is referred to as the
\textit{constraint set}. We then define the exact
\textit{feasibility set} $\Theta_i$ as the intersection of the
constraint sets over the time instants $l=1, \ldots, i$, which is
given by
\begin{equation}\label{8}
\Theta_i={\bigcap_{l=1}^{i}}_{(s_0, \boldsymbol x)\in{\boldsymbol
S}}\mathcal H_l,
\end{equation}
where $s_0$ is the SOI and $\boldsymbol S$ is the set including all
possible data pairs $\{s_0, \boldsymbol x\}$. The aim of (\ref{8})
is to develop adaptive algorithms that update the parameters such
that they will always remain within the feasibility set. In theory,
$\Theta$ should encompass all the pairs of solutions $\{\boldsymbol
T_r, \bar{\boldsymbol w}\}$ that satisfy the bounded constraint
until $i\rightarrow\infty$. In practice, $\boldsymbol S$ cannot be
traversed all over. It implies that a larger space of the data pairs
provided by the observations leads to a smaller feasibility set.
Thus, as the number of data pairs (or ``$i$") increases, there are
fewer pairs of $\{\boldsymbol T_r, \bar{\boldsymbol w}\}$ that can
be found to satisfy the constraint. Under this condition, we define
the \textit{membership set} $\Psi_i=\bigcap_{l=1}^{i}\mathcal H_l$
as the practical set of the proposed JIO-SM scheme. It is obvious
that $\Theta$ is a limiting set of $\Psi$. These two sets will be
equal if the data pairs traverse $\boldsymbol S$ completely.

The proposed JIO-SM framework introduces the principle of the SMF
technique into the constrained reduced-rank signal processing for
reducing the computational complexity.
The reduced number of parameters and data-selective
updates reduce the complexity. It should be remarked that,
due to the time-varying nature of many practical environments, the
time-varying bound should be selected appropriately to account for the
characteristics of the environment. Moreover, the use of an appropriate bound will lead
to highly effective variable step-sizes and forgetting factors for
the SG-based and RLS-based algorithms, respectively, an increased
convergence speed and improved tracking ability. We will detail their relations next.

\section{Proposed JIO-SM Adaptive Algorithms}

We derive SG-based and RLS-based adaptive algorithms for the
proposed JIO-SM scheme. They are developed according to the
reduced-rank LCMV optimization problem that incorporates the
time-varying bounded constraint on the amplitude of the array
output. The problem is defined as:
\begin{equation}\label{9}
\begin{split}
&\textrm{minimize}~~~\mathbb E[|\bar{\boldsymbol w}^H\boldsymbol
T_r^H\boldsymbol x|^2]=\bar{\boldsymbol
w}^H\bar{\boldsymbol R}\bar{\boldsymbol w}\\
&\textrm{subject~to}~~~\bar{\boldsymbol w}^H\bar{\boldsymbol
a}(\theta_0)=\gamma~\textrm{and}~|\bar{\boldsymbol w}^H\boldsymbol
T_r^H\boldsymbol x|^2=\delta^2,
\end{split}
\end{equation}

The optimization problem in (\ref{9}) is a function of $\boldsymbol
T_r$ and $\bar{\boldsymbol w}$. {  In order to obtain a solution, we
employ an alternating optimization strategy, which is equivalent to
fixing $\bar{\boldsymbol w}$ and computing $\boldsymbol T_r$ with a
suitable adaptive algorithm followed by another step with
$\boldsymbol T_r$ fixed and the use of another adaptive algorithm to
adjust $\bar{\boldsymbol w}$. This will be pursued in what follows
with constrained SG and RLS-type algorithms for which a time-varying
bound $\delta[i]$ determines a set of solutions $\{\boldsymbol T_r,
\bar{\boldsymbol w}\}$ within the constraint set $\mathcal H$ at
each time instant. Regarding the convergence of this type of
strategy, a general alternating optimization strategy has been shown
in \cite{niesen} to converge to the global minimum. In our studies,
problems with local minima have not been found although a proof of
convergence is left for future work.}

\subsection{Proposed JIO-SM-SG Algorithm}
In order to solve the optimization problem by the SG-based adaptive
algorithm, we employ the Lagrange multiplier method \cite{Haykin} to
transform the constrained problem into an unconstrained one, which
is
\begin{equation}\label{10}
J(\boldsymbol T_r, \bar{\boldsymbol w})=\mathbb
E\big[\bar{\boldsymbol w}^H\boldsymbol T_r^H\boldsymbol x\boldsymbol
x^H\boldsymbol T_r\bar{\boldsymbol
w}\big]+2\eta\mathfrak{R}\big[\bar{\boldsymbol w}^H\bar{\boldsymbol
a}(\theta_0)-\gamma\big],
\end{equation}
where $\eta$ is the Lagrange multiplier and $\mathfrak R[\cdot]$
selects the real part of the quantity. It should be remarked that
the bounded constraint $|\bar{\boldsymbol w}^H\boldsymbol
T_r^H\boldsymbol x|^2=\delta^2$ is not included in (\ref{10}). This
is because a point estimate can be obtained from (\ref{10}) whereas
the bounded constraint determines a set of $\{\boldsymbol T_r,
\bar{\boldsymbol w}\}$ (also including the solution from
(\ref{10})). We use the constraint on the steering vector of the SOI
to obtain a solution and employ the constraint to expand it to a
hyperplane (multiple solutions).

Assuming $\bar{\boldsymbol w}$ is known, taking the instantaneous
gradient of (\ref{10}) with respect to $\boldsymbol T_r$, equating
it to a zero matrix and solving for $\eta$, we have
\begin{equation}\label{11}
\boldsymbol T_r(i+1)=\boldsymbol
T_r(i)-\mu_Ty^{\ast}(i)\big[\boldsymbol I-\frac{\boldsymbol
a(\theta_0)\boldsymbol a^H(\theta_0)}{{\boldsymbol
a}^H(\theta_0){\boldsymbol a}(\theta_0)}\big]\boldsymbol
x(i)\bar{\boldsymbol w}^H(i),
\end{equation}
where $\mu_T$ is the step size value for the update of the
transformation matrix and $\boldsymbol I$ is the corresponding
identity matrix. Note that we use the adaptive version to perform
parameter estimation and thus we include ``$i$" in the related
quantities.

Assuming $\boldsymbol T_r$ is known, computing the instantaneous
gradient of (\ref{10}) with respect to $\bar{\boldsymbol w}$,
equating it a null vector and solving for $\eta$, we obtain
\begin{equation}\label{12}
\bar{\boldsymbol w}(i+1)=\bar{\boldsymbol
w}(i)-\mu_{\bar{w}}y^{\ast}(i)\big[\boldsymbol
I-\frac{\bar{\boldsymbol a}(\theta_0)\bar{\boldsymbol
a}^H(\theta_0)}{\bar{\boldsymbol a}^H(\theta_0)\bar{\boldsymbol
a}(\theta_0)}\big]\bar{\boldsymbol x}(i),
\end{equation}
where $\mu_{\bar{w}}$ is the step size value for the update of the
reduced-rank weight vector.

The SMF technique provides an effective way to adjust the step size
values and to improve the performance. SMF algorithms with the
predetermined bounds were reported in \cite{Nagaraj}. However, a
predetermined bound always has the risk of underbounding (the bound
is smaller than the actual one) or overbounding (the bound is larger
than the actual one).
Instead of the predetermined bound, we use a time-varying bound in
the proposed JIO-SM-SG algorithm to adjust the step size values for
offering a good tradeoff between the
convergence and the misadjustment, which are  
\begin{equation}\label{13}
\mu_{T}(i)=\left\{ \begin{array}{ccc}
                  \frac{1-\frac{\delta(i)}{|y(i)|}}{\bar{\boldsymbol w}^H(i)\bar{\boldsymbol w}(i)\boldsymbol x^H(i)[\boldsymbol I-\boldsymbol a(\theta_0)\boldsymbol a^H(\theta_0)]{\boldsymbol x}(i)} & \textrm {if}|y(i)|^2\geq\delta^2(i)\\
                  0 & \textrm {otherwise}\\
                  \end{array}\right.
\end{equation}
and
\begin{equation}\label{14} \mu_{\bar{w}}(i)=\left\{
\begin{array}{ccc}
                 \frac{1-\frac{\delta(i)}{|y(i)|}}{\bar{\boldsymbol x}^H(i)[\boldsymbol I-\frac{\bar{\boldsymbol a}(\theta_0)\bar{\boldsymbol a}^H(\theta_0)}{\bar{\boldsymbol a}^H(\theta_0)\bar{\boldsymbol a}(\theta_0)}]\bar{\boldsymbol
                 x}(i)} & \textrm {if}~|y(i)|^2\geq\delta^2(i)\\
                 0 & \textrm{otherwise},\\
                 \end{array}\right.
\end{equation}
where the derivations are provided in the appendix.

The proposed JIO-SM-SG algorithm consists of the equations
(\ref{11})-(\ref{14}), where the expression of the time-varying
bound $\delta(i)$ will be addressed later in this section. From
(\ref{11}) and (\ref{12}), the transformation matrix and the
reduced-rank filter depend on each other, which provides a joint
iterative exchange to utilize the key information of the
reduced-rank received vector more effectively, and thus leads to an
improved performance. The SMF technique with the time-varying bound
is employed to determine a set of estimates $\{\boldsymbol T_r(i),
\bar{\boldsymbol w}(i)\}$ that satisfy the bounded constraint
(constraint set $\mathcal H_i$). The computational complexity is
reduced significantly due to the data-selective updates. The
proposed algorithm is more robust to dynamic scenarios compared to
their SG-based counterparts.

\subsection{Proposed JIO-SM-RLS Algorithm}

The constrained optimization problem in (\ref{9}) can be transformed
into an unconstrained least squares (LS) one by the Lagrange
multipliers. The Lagrangian is given by {\small
\begin{equation}\label{15}
\begin{split}
&J(\boldsymbol T_r, \bar{\boldsymbol
w})=\sum_{l=1}^{i-1}\lambda_1^{i-l}(i)\bar{\boldsymbol
w}^H(i)\boldsymbol T_r^H(i)\boldsymbol x(l)\boldsymbol
x^H(l)\boldsymbol T_r(i)\bar{\boldsymbol w}(i)\\
&+2\lambda_1(i)\mathfrak{R}\big[|\bar{\boldsymbol w}^H(i)\boldsymbol
T_r^H(i)\boldsymbol
x(i)|^2-\delta^2(i)\big]+2\lambda_2\mathfrak{R}\big[\bar{\boldsymbol
w}^H(i)\bar{\boldsymbol a}(\theta_0)-\gamma\big],
\end{split}
\end{equation}}
where $\lambda_1(i)$ plays the role of the forgetting factor and the
Lagrange multiplier with respect to the bounded constraint.
This coefficient is helpful to estimate the received covariance matrix
in a recursive form and utilize the matrix inversion lemma.
The coefficient $\lambda_2$ is another Lagrange multiplier for
the constraint on the steering vector of the SOI.

Assuming $\bar{\boldsymbol w}(i-1)$ is known, taking the gradient of
$\boldsymbol T_r(i)$ with respect to (\ref{15}) and employing the
matrix inversion lemma \cite{Haykin}, we have
\begin{equation}\label{16}
\boldsymbol T_r(i)=\frac{\gamma\boldsymbol P(i)\boldsymbol
a(\theta_0)}{\boldsymbol a^H(\theta_0)\boldsymbol P(i)\boldsymbol
a(\theta_0)}\frac{\bar{\boldsymbol w}^H(i-1)}{\|\bar{\boldsymbol
w}(i-1)\|^2},
\end{equation}
where $\boldsymbol R(i)=\boldsymbol R(i-1)+\lambda_1(i)\boldsymbol
x(i)\boldsymbol x^H(i)$ (note that this expression is given under an
assumption that $\lambda_1(i)$ is close to $1$ in order to make it
according with the setting of the forgetting factor \cite{Haykin},
so as $\bar{\boldsymbol R}(i)$ in the following) and $\boldsymbol
P(i)=\boldsymbol R^{-1}(i)$ is calculated in a recursive form
\begin{equation}\label{17}
\boldsymbol k(i)=\frac{\boldsymbol P(i-1)\boldsymbol
x(i)}{1+\lambda_1(i)\boldsymbol x^H(i)\boldsymbol P(i-1)\boldsymbol
x(i)}
\end{equation}
\begin{equation}\label{18}
\boldsymbol P(i)=\boldsymbol P(i-1)-\lambda_1(i)\boldsymbol
k(i)\boldsymbol x^H(i)\boldsymbol P(i-1).
\end{equation}
The derivation of (\ref{16}) is given in the appendix.

Given the assumptions that $\boldsymbol T_r(i)$ is known and
$\lambda_1(i)\rightarrow1$, computing the gradient of
$\bar{\boldsymbol w}(i)$ with respect to (\ref{15}), we get
\begin{equation}\label{19}
\bar{\boldsymbol w}(i)=\frac{\gamma\bar{\boldsymbol
P}(i)\bar{\boldsymbol a}(\theta_0)}{\bar{\boldsymbol
a}^H(\theta_0)\bar{\boldsymbol P}(i)\bar{\boldsymbol a}(\theta_0)},
\end{equation}
where $\bar{\boldsymbol R}(i)=\bar{\boldsymbol
R}(i-1)+\lambda_1(i)\bar{\boldsymbol x}(i)\bar{\boldsymbol x}^H(i)$
and $\bar{\boldsymbol P}(i)=\bar{\boldsymbol R}^{-1}(i)$ is
calculated by
\begin{equation}\label{20}
\bar{\boldsymbol k}(i)=\frac{\bar{\boldsymbol
P}(i-1)\bar{\boldsymbol x}(i)}{1+\lambda_1(i)\bar{\boldsymbol
x}^H(i)\bar{\boldsymbol P}(i-1)\bar{\boldsymbol x}(i)}
\end{equation}
\begin{equation}\label{21}
\bar{\boldsymbol P}(i)=\bar{\boldsymbol
P}(i-1)-\lambda_1(i)\bar{\boldsymbol k}(i)\bar{\boldsymbol
x}^H(i)\bar{\boldsymbol P}(i-1).
\end{equation}

The coefficient $\lambda_1(i)$ is important to the updates of
$\boldsymbol T_r(i)$ and $\bar{\boldsymbol w}(i)$. In order to
obtain its expression, we substitute (\ref{16}) and (\ref{19}) into
the constraint in (\ref{9}), which leads to  {\small
\begin{equation}\label{22}
\begin{split}
\lambda_1(i)= \left\{ \begin{array}{ccc}
                  \frac{\boldsymbol a^H(\theta_0)\boldsymbol
P(i-1)[\delta(i)\boldsymbol a(\theta_0)-\gamma^2\boldsymbol
x(i)]}{\boldsymbol a^H(\theta_0)\boldsymbol k(i)\boldsymbol
x^H(i)\boldsymbol P(i-1)[\delta(i)\boldsymbol
a(\theta_0)-\gamma^2\boldsymbol x(i)]}
                    & \textrm{if} ~|y(i)|^2\geq\delta^2(i)\\
                  0 & \textrm {otherwise},\\
                  \end{array}\right.
\end{split}
\end{equation}
} It is clear that $\lambda_1(i)$ involves the time-varying bound,
the full-rank received vector and the related quantities. It
provides a way to track the changes of $\delta(i)$ and control the
weighting of $\boldsymbol P(i)$. The proposed JIO-SM-RLS algorithm
corresponds to equations (\ref{16})-(\ref{22}), where $\rho$ and
$\varrho$ are small positive values for regularization, and
$\boldsymbol T_r(0)$ and $\bar{\boldsymbol w}(0)$ are used for
initialization. The joint iterative exchange of information between
$\boldsymbol T_r(i)$ and $\bar{\boldsymbol w}(i)$ is achieved from
their update equations. The coefficient $\lambda_1(i)$ is calculated
only if the constraint cannot be satisfied, so as the parameters'
update. All the pairs of $\{\boldsymbol T_r(i), \bar{\boldsymbol
w}(i)\}$ ensuring the bounded constraint until time instant $i$ are
in the feasibility set $\Theta_i$. The proposed JIO-SM-RLS algorithm
has better performance and lower computational cost than the
existing reduced-rank algorithms.

\subsection{Time-varying Bound}

The time-varying bound $\delta(i)$ is a single coefficient to check
if the parameter update is carried out or not. In other words, it is
an important criterion to measure the quality of the parameters that
could be included in the feasibility set $\Theta_i$. Besides, it is
better if $\delta(i)$ could reflect the characteristics
(time-varying nature) of the environment since it benefits the
estimation and the tracking of the proposed algorithms. From
(\ref{7}) and (\ref{8}), $\delta(i)$ cannot be chosen too stringent
for avoiding an empty $\Theta_i$ with respect to a given model space
of interest. Here, we introduce a parameter dependent bound (PDB)
that is similar to the work reported in \cite{Guo2} but which
considers both $\boldsymbol T_r(i)$ and $\bar{\boldsymbol w}(i)$.
The proposed time-varying bound is
\begin{equation}\label{23}
\delta(i)=\beta\delta(i-1)+(1-\beta)\sqrt{\alpha\|\boldsymbol
T_r(i)\bar{\boldsymbol w}(i)\|^2\hat{\sigma}_n^2(i)},
\end{equation}
where $\beta$ is a positive value close to $1$ ($\beta=0.99$ in general), which is set to guarantee
an proper time-averaged estimate of the evolutions of the weight
vector $\boldsymbol w(i)$, $\alpha$($\alpha>1$) is a tuning
coefficient that impacts the update rate and the convergence, and
$\hat{\sigma}_n^2(i)$ is an estimate of the noise power, which is
assumed to be known at the receiver. The term $\|\boldsymbol
T_r(i)\bar{\boldsymbol w}(i)\|^2\hat{\sigma}_n^2(i)$ is the variance
of the inner product of the weight vector with the noise that
provides information on the evolution of $\boldsymbol T_r(i)$ and
$\bar{\boldsymbol w}(i)$. It formulates a relation between the
estimated parameters and the environmental coefficients. This kind
of update provides a smoother evolution of the weight vector
trajectory and thus avoids too high or low values of the squared
norm of the weight vector. As $\delta(i)$ is chosen properly, it
ensures that the feasibility set $\Theta_i$ is nonempty and any
point in it is a valid estimate with respect to the constraint set
$\mathcal H_i$.

\section{Analysis}
In this section, we give a complexity analysis of the proposed
algorithms and compare them with the existing algorithms. An MSE
expression to predict the performance of the proposed JIO-SM-SG
algorithm is derived. We also give the stability analysis and study
the properties of the optimization problem.

\subsection{Complexity Analysis}

In \cite{Wang}, the computational complexity required for the
existing full-rank and reduced-rank adaptive algorithms for each
time instant (snapshot) is reported. Here, due to the data-selective
updates, we calculate the complexity for the whole number of
snapshots $N$ to provide a fair comparison. The computational cost
is measured in terms of the number of complex arithmetic operations,
i.e., additions and multiplications. The results are listed in Table
\ref{tab: Complexity}, where $r$ is the number of rank, $m$ is the
number of sensor elements, $N$ is the number of snapshots, and
$\tau$ ($0<\tau\leq1$) is the update rate for the adaptive
algorithms with the SMF technique, which is obtained by finding the
number of updates for a fixed $N$.

\begin{table*}[ht]
\centering \caption{\normalsize Computational complexity of
algorithms} \footnotesize \label{tab: Complexity}
\begin{tabular}{l c c}
\hline
Algorithm & Additions                   & Multiplications\\
\hline
FR-SG \cite{Frost}    & $N(3m-1)$                        & $N(4m+1)$ \\
FR-SM-SG \cite{Diniz}  & $2Nm+3\tau Nm$              & $N(2m+5)+\tau N(4m+3)$\\
FR-RLS \cite{Haykin}   & $N(4m^2-m-1)$            & $N(5m^2+5m-1)$\\
FR-SM-RLS \cite{Lamare4}  & $2Nm+\tau N(4m^2-1)$   & $N(2m+5)+\tau N(5m^2+6m+2)$\\
MSWF-SG \cite{Goldstein}        & $N(rm^2+(r+1)m+2r-2)$             & $N(rm^{2}+2rm+5r+2)$ \\
MSWF-RLS \cite{Honig2}   & $N(rm^2+(r+1)m+4r^2-3r-1)$   & $N((r+1)m^2+2rm+5r^2+4r)$\\
AVF \cite{Pados}         & $N((4r+5)m^2+(r-1)m-2r-1)$   & $N((5r+8)m^2+(3r+2)m)$\\
JIO-SG \cite{jio_lcmv_esp}       & $N(4rm+m+2r-3)$            & $N(4rm+m+7r+3)$\\
JIO-SM-SG  &  $2Nrm+\tau N(3rm+2m+2r-4)$    & $N(2rm+m+r+5)+\tau N(3rm+2m+8r+7)$\\
JIO-RLS \cite{Wang}   & $N(4m^2+(2r-1)m+4r^2-4r-1)$           & $N(5m^2+(3r+3)m+6r^2+4r)$\\
JIO-SM-RLS   & $2Nmr+\tau N(4m^2+rm+m+4r^2-6r-1)$    & $N(2rm+m+r+5)+\tau N(5m^2+2rm+5m+6r^2+3r+3)$\\
\hline
\end{tabular}
\end{table*}

From Table \ref{tab: Complexity}, we find that the complexity of the
existing and proposed algorithms depends more on $N$ and $m$
(especially for large arrays) since they are much larger than $r$,
which is often selected around a small range. The value of the
update rate $\tau$ impacts the complexity significantly.
Specifically, for a small value of $\tau$, the complexity of the
algorithms with the SMF technique is much lower than their
counterparts with $100\%$ updates since the parameter estimation
procedures only perform with a small number of snapshots. For a very
large $\tau$ (e.g., $\tau=1$), the SM-based algorithm is a little
more complex than their counterparts due to the calculations of the
time-varying bound, step size values (for the SG-based algorithms),
and the forgetting factor (for the RLS-based algorithms). In most
cases, it only needs a small number of updates to achieve parameter
estimation and thus reduces the computational cost.

\begin{figure}[htb]
\begin{minipage}[h]{1.0\linewidth}
  \centering
  \centerline{\epsfig{figure=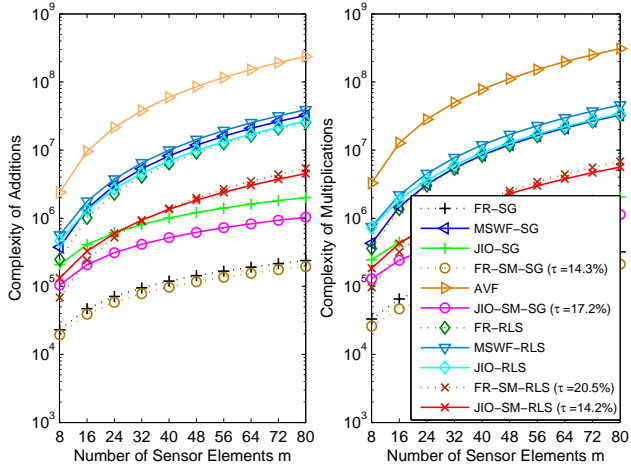,scale=0.6}} \vspace{-0.5em}\caption{Complexity
  in terms of arithmetic operations versus the number of sensor elements $m$.} \label{fig:complexity}
\end{minipage}
\end{figure}

Fig. \ref{fig:complexity} provides a more direct way to illustrate
the complexity requirements for the algorithms compared. It shows the
complexity in terms of additions and multiplications versus the
number of sensor elements $m$. Note that the values of $r$ and
$\tau$ are different with respect to different algorithms, which are
set to make a good tradeoff between the output performance and the
complexity. Their specific values are given in the figure. It is
clear that the reduced-rank adaptive algorithms are more complex
than the full-rank ones due to the generation of the transformation
matrix. The adaptive algorithms with the SMF technique save the
computational cost significantly. The proposed JIO-SM-SG and
JIO-SM-RLS algorithms have a complexity slightly higher than their
full-rank algorithms but much lower than the existing reduced-rank
methods. As $N$ or/and $m$ increase, this advantage is more obvious.
It is worth mentioning that the complexity reduction due to the
data-selective updates does not degrade the performance. This
will be shown in the simulation results.

\subsection{Stability Analysis}

In order to establish conditions for the stability of the proposed
JIO-SM-SG algorithm, we define $\boldsymbol e_{T}(i)=\boldsymbol
T_r(i)-\boldsymbol T_{r,\textrm{opt}}$ and $\boldsymbol
e_{\bar{w}}(i)=\bar{\boldsymbol w}(i)-\bar{\boldsymbol
w}_{\textrm{opt}}$ with $\boldsymbol T_{r,\textrm{opt}}$ and
$\bar{\boldsymbol w}_{\textrm{opt}}$ being the optimal solutions of
the transformation matrix and the reduced-rank filter, respectively. The expression
 of $\boldsymbol T_{r,\textrm{opt}}$ can be obtained by taking the gradient of \eqref{10} with respect to $\boldsymbol T_r$, i.e., $\boldsymbol T_{r,\textrm{opt}}=\frac{\gamma\boldsymbol R\boldsymbol a(\theta_0)}{\boldsymbol a^H(\theta_0)\boldsymbol R\boldsymbol a(\theta_0)}\frac{\bar{\boldsymbol w}_{\textrm{opt}}^H}{\|\bar{\boldsymbol w}_{\textrm{opt}}\|^2}$, where $\bar{\boldsymbol w}_{\textrm{opt}}$ has been given in \eqref{6}.
By substituting (\ref{11}) and (\ref{12}) into $\boldsymbol
e_{T}(i)$ and $\boldsymbol e_{\bar{w}}(i)$, respectively, and
rearranging the terms, we have
\begin{equation}\label{24}
\begin{split}
\boldsymbol e_{T}(i+1)=\boldsymbol T_1(i)\boldsymbol
e_T(i)-\mu_{T}(i)\boldsymbol T_2(i)+\mu_T(i)\boldsymbol T_3(i),
\end{split}
\end{equation}
\begin{equation}\label{25}
\boldsymbol e_{\bar{w}}(i+1)=\bar{\boldsymbol W}_1(i)\boldsymbol
e_{\bar{w}}(i)-\mu_{\bar{w}}(i)\bar{\boldsymbol W}_2(i),
\end{equation}
where\\
$\boldsymbol T_1(i)=\boldsymbol I-\mu_{T}(i)[\boldsymbol
I-\frac{\boldsymbol a(\theta_0)\boldsymbol
a^H(\theta_0)}{{\boldsymbol a}^H(\theta_0){\boldsymbol
a}(\theta_0)}]\boldsymbol x(i)\boldsymbol
x^H(i)$;\\
$\boldsymbol T_2(i)=[\boldsymbol I-\frac{\boldsymbol
a(\theta_0)\boldsymbol a^H(\theta_0)}{{\boldsymbol
a}^H(\theta_0){\boldsymbol a}(\theta_0)}]\boldsymbol x(i)\boldsymbol
x^H(i)\bar{\boldsymbol w}^H(i)\boldsymbol e_{\bar{w}}(i)\boldsymbol
T_r(i)$;\\
$\boldsymbol T_3(i)=[\boldsymbol I-\frac{\boldsymbol
a(\theta_0)\boldsymbol a^H(\theta_0)}{{\boldsymbol
a}^H(\theta_0){\boldsymbol a}(\theta_0)}]\boldsymbol x(i)\boldsymbol
x^H(i)[(1-\bar{\boldsymbol w}^H(i)\bar{\boldsymbol
w}_{\textrm{opt}})\boldsymbol T_r(i)-\boldsymbol
T_{r,\textrm{opt}}]$;\\
$\bar{\boldsymbol W}_1(i)=\boldsymbol I-\mu_{\bar{w}}(i)[\boldsymbol
I-\frac{\bar{\boldsymbol a}(\theta_0)\bar{\boldsymbol
a}^H(\theta_0)}{\bar{\boldsymbol a}^H(\theta_0)\bar{\boldsymbol
a}(\theta_0)}]\bar{\boldsymbol x}(i)\bar{\boldsymbol x}^H(i)$;\\
$\bar{\boldsymbol W}_2(i)=[\boldsymbol I-\frac{\bar{\boldsymbol
a}(\theta_0)\bar{\boldsymbol a}^H(\theta_0)}{\bar{\boldsymbol
a}^H(\theta_0)\bar{\boldsymbol a}(\theta_0)}]\bar{\boldsymbol
x}(i)\bar{\boldsymbol x}^H(i)\bar{\boldsymbol w}_{\textrm{opt}}$.

Since we are dealing with a joint optimization procedure, both the
transformation matrix and the reduced-rank filter have to be
considered jointly. Besides, the time-varying bound should be
investigated. By substituting \eqref{13} and \eqref{14} into
\eqref{24} and \eqref{25}, respectively, and taking expectations, we
get
\begin{equation}\label{26}
\begin{bmatrix}
\mathbb E[\boldsymbol e_{T}(i+1)]\\
\mathbb E[\boldsymbol e_{\bar{w}}(i+1)]\\
\end{bmatrix}=\begin{bmatrix}
\boldsymbol U_1 & \boldsymbol 0\\
\boldsymbol 0 & \bar{\boldsymbol U}_2\\\end{bmatrix}\begin{bmatrix} \mathbb E[\boldsymbol e_{T}(i)]\\
\mathbb E[\boldsymbol e_{\bar{w}}(i)]\\
\end{bmatrix}+\begin{bmatrix}
\boldsymbol V_1\\
\bar{\boldsymbol V}_2\\\end{bmatrix},
\end{equation}
where\\
$\boldsymbol U_1=\boldsymbol
I-\frac{\big(1-\frac{\delta(i)}{|y(i)|}\big)[\boldsymbol
I-\frac{\boldsymbol a(\theta_0)\boldsymbol
a^H(\theta_0)}{{\boldsymbol a}^H(\theta_0){\boldsymbol
a}(\theta_0)}]\boldsymbol x(i)\boldsymbol x^H(i)}{\|\bar{\boldsymbol
w}(i)\|^2\boldsymbol x^H(i)[\boldsymbol I-\frac{\boldsymbol
a(\theta_0)\boldsymbol a^H(\theta_0)}{{\boldsymbol
a}^H(\theta_0){\boldsymbol a}(\theta_0)}]\boldsymbol x(i)}$;\\
$\bar{\boldsymbol U}_2=\boldsymbol
I-\frac{\big(1-\frac{\delta(i)}{|y(i)|}\big)[\boldsymbol
I-\frac{\bar{\boldsymbol a}(\theta_0)\bar{\boldsymbol
a}^H(\theta_0)}{\bar{\boldsymbol a}^H(\theta_0)\bar{\boldsymbol
a}(\theta_0)}]\bar{\boldsymbol x}(i)\bar{\boldsymbol
x}^H(i)}{\bar{\boldsymbol x}^H(i)[\boldsymbol
I-\frac{\bar{\boldsymbol a}(\theta_0)\bar{\boldsymbol
a}^H(\theta_0)}{\bar{\boldsymbol a}^H(\theta_0)\bar{\boldsymbol
a}(\theta_0)}]\bar{\boldsymbol
x}(i)}$;\\
${\boldsymbol
V}_1=\frac{\big(1-\frac{\delta(i)}{|y(i)|}\big)[-\boldsymbol
T_2(i)+\boldsymbol T_3(i)]}{\|\bar{\boldsymbol w}(i)\|^2\boldsymbol
x^H(i)[\boldsymbol I-\frac{\boldsymbol a(\theta_0)\boldsymbol
a^H(\theta_0)}{{\boldsymbol
a}^H(\theta_0){\boldsymbol a}(\theta_0)}]\boldsymbol x(i)}$;\\
$\bar{\boldsymbol
V}_2=-\frac{\big(1-\frac{\delta(i)}{|y(i)|}\big)\bar{\boldsymbol
W}_2(i)}{\bar{\boldsymbol x}^H(i)[\boldsymbol
I-\frac{\bar{\boldsymbol a}(\theta_0)\bar{\boldsymbol
a}^H(\theta_0)}{\bar{\boldsymbol a}^H(\theta_0)\bar{\boldsymbol
a}(\theta_0)}]\bar{\boldsymbol x}(i)}$.

From (\ref{26}), it implies that the stability of the proposed
JIO-SM-SG algorithm depends on the spectral radius of $\boldsymbol
U=\textrm{diag}[\boldsymbol U_1, \bar{\boldsymbol U}_2]$. The step
size values should satisfy the condition that the eigenvalues of
$\boldsymbol U^H\boldsymbol U$ are less than one for convergence.
The variable step size values calculated by \eqref{13} and
\eqref{14} follow this condition with the
bounded constraint. Unlike the stability analysis of existing
adaptive algorithms, the terms in the proposed algorithm are more
involved and depend on each other as evidenced by the equations in
$\boldsymbol U$, $\boldsymbol V_1$ and $\bar{\boldsymbol V}_2$.

\subsection{Prediction of The Trend of MSE}
In this part, we derive expressions to predict the trend of the MSE
for the proposed JIO-SM-SG algorithm. The following analysis begins
with the conventional MSE analysis of \cite{Haykin} and then
involves the novel parameters $\boldsymbol T_r(i)$ and
$\bar{\boldsymbol w}(i)$ due to the joint optimization property. The
data-selective updates of the SMF technique is also considered in
the analysis by introducing a new coefficient $P_e(i)$ in the update
equations.

Let us define the the estimation error at time instant $i$ to be
\begin{equation}\label{27}
e(i)=d_0(i)-y(i)=e_0(i)-\boldsymbol e_w^H(i)\boldsymbol x(i),
\end{equation}
where $d_0(i)$ denotes the transmitted data of the desired user,
$\boldsymbol e_w(i)=\boldsymbol w(i)-\boldsymbol w_{\textrm{opt}}$
with $\boldsymbol w_{\textrm{opt}}$ being the optimal weight
solution, and $e_0(i)=d_0(i)-\boldsymbol
w_{\textrm{opt}}^H\boldsymbol x(i)$. The filter $\boldsymbol
w(i)=\boldsymbol T_r(i)\bar{\boldsymbol w}(i)$ with $m$ parameters
is the $r$-rank approximation of a full-rank filter obtained with an
inverse transformation \cite{jio_lcmv_esp} processed by $\boldsymbol
T_r(i)$.

The MSE following the time instant $i$ is given by
\begin{equation}\label{28}
\begin{split}
J_{\textrm{mse}}(i)&=\mathbb E[|e(i)|^2]\\
&=J_{\textrm{min}}+\mathbb
E[\boldsymbol e_w^H(i)\boldsymbol R\boldsymbol
e_w(i)]\\
&=J_{\textrm{min}}+\sigma_x^2\textrm{tr}\{\textrm{cov}[\boldsymbol
e_w(i)]\}
\end{split}
\end{equation}
where $J_{\textrm{min}}$ is the minimum MSE (MMSE) produced by the
optimal LCMV solution and
$\sigma_x^2\textrm{tr}\{\textrm{cov}[\boldsymbol e_w(i)]\}$ denotes
the excess MSE (EMSE) with $\sigma_x^2=B_0^2
b_0^2+B_1^2b_1^2+\ldots+B_{q-1}^2b_{q-1}^2$ being the summed
variance of the transmitted data and $B_k$ ($k=0, \ldots, q-1$)
being the amplitude. Assuming $d_k$ is independent and identically
distributed (i.i.d.), the MMSE can be expressed by
\begin{equation}\label{29}
\begin{split}
J_{\textrm{min}}&=\mathbb E[|e_0(i)|^2]\\
&=\mathbb
E[|d_0(i)|^2]-\boldsymbol w_{\textrm{opt}}^H\boldsymbol
a(\theta_0)-\boldsymbol a^H(\theta_0)\boldsymbol
w_{\textrm{opt}}+\boldsymbol w_{\textrm{opt}}^H\boldsymbol
R\boldsymbol w_{\textrm{opt}}.
\end{split}
\end{equation}
Considering the inverse transformation, the weight error vector
becomes,
\begin{equation}\label{30}
\begin{split}
\boldsymbol e_w(i)&=\boldsymbol T_r(i)\bar{\boldsymbol w}(i)-\boldsymbol T_{r,\textrm{opt}}\bar{\boldsymbol w}_{\textrm{opt}}\\
&=\boldsymbol e_{T}(i)\boldsymbol e_{\bar{w}}(i)+\boldsymbol
T_{r,\textrm{opt}}\boldsymbol e_{\bar{w}}(i)+\boldsymbol
e_{T}(i)\bar{\boldsymbol w}_{\textrm{opt}}.
\end{split}
\end{equation}
To provide further analysis, we use (\ref{11}) and \eqref{12} and
consider the data-selective updates of the SMF technique given by
{\small
\begin{equation}\label{31}
\boldsymbol T_r(i+1)=\boldsymbol
T_r(i)-P_e(i)\mu_T(i)y^{\ast}(i)\big[\boldsymbol I-\frac{\boldsymbol
a(\theta_0)\boldsymbol a^H(\theta_0)}{{\boldsymbol
a}^H(\theta_0){\boldsymbol a}(\theta_0)}\big]\boldsymbol
x(i)\bar{\boldsymbol w}^H(i),
\end{equation}}
\begin{equation}\label{32}
\bar{\boldsymbol w}(i+1)=\bar{\boldsymbol
w}(i)-P_e(i)\mu_{\bar{w}}(i)y^{\ast}(i)\big[\boldsymbol
I-\frac{\bar{\boldsymbol a}(\theta_0)\bar{\boldsymbol
a}^H(\theta_0)}{\bar{\boldsymbol a}^H(\theta_0)\bar{\boldsymbol
a}(\theta_0)}\big]\bar{\boldsymbol x}(i),
\end{equation}
where $\mu_T(i)$ and $\mu_{\bar{w}}(i)$ are the variable step size
values following the time-varying bound, and $P_e(i)$ is a
coefficient modeling the probability of updating the filter
parameters with respect to a given time instant $i$, namely,
$P_e(i)=P[|y(i)|^2>\delta^2(i)]$. Note that $P_e(i)$ is same for
(\ref{31}) and \eqref{32} since $\boldsymbol T_r(i)$ and
$\bar{\boldsymbol w}(i)$ depend on each other and update jointly.

Substituting \eqref{31} and \eqref{32} into \eqref{30} and making
some rearrangements, we have
\begin{equation}\label{33}
\begin{split}
\boldsymbol e_{w}(i+1)=&\boldsymbol
e_w(i)-P_e(i)\mu_{T}(i)y^{\ast}(i)\boldsymbol G_r(i)\bar{\boldsymbol
w}(i)\\
&-P_e(i)\mu_{\bar{w}}(i)y^{\ast}(i)\boldsymbol
T_r(i)\bar{\boldsymbol
g}(i)\\
&+P_e^2(i)\mu_{T}(i)\mu_{\bar{w}}(i)[y^{\ast}(i)]^2\boldsymbol
G_r(i)\boldsymbol g(i),
\end{split}
\end{equation}
where {$\boldsymbol G_r(i)=\big[\boldsymbol I-\boldsymbol
a(\theta_0)\boldsymbol a^H(\theta_0)\big]\boldsymbol
x(i)\bar{\boldsymbol w}^H(i)$}, and $\bar{\boldsymbol
g}(i)=\big[\boldsymbol I-\frac{\bar{\boldsymbol
a}(\theta_0)\bar{\boldsymbol a}^H(\theta_0)}{\bar{\boldsymbol
a}^H(\theta_0)\bar{\boldsymbol a}(\theta_0)}\big]\bar{\boldsymbol
x}(i)$. By using \eqref{13} and \eqref{14} and applying
$y(i)=\bar{\boldsymbol w}^H(i)\bar{\boldsymbol x}(i)$ to \eqref{30},
we get
\begin{equation}\label{34}
\begin{split}
\boldsymbol e_w(i+1)=[\boldsymbol I-\boldsymbol V(i)]\boldsymbol
e_w(i)-\boldsymbol V(i)\boldsymbol
w_{\textrm{opt}}+\tau(i)\boldsymbol G_r(i)\bar{\boldsymbol g}(i),
\end{split}
\end{equation}
where\\
$\boldsymbol
V(i)=P_e(i)[1-\frac{\delta(i)}{|y(i)|}][\frac{\boldsymbol
G_r(i)\bar{\boldsymbol w}(i)\boldsymbol x^H(i)}{\bar{\boldsymbol
w}^H(i)\boldsymbol G_r^H(i)\boldsymbol x(i)}+\frac{\boldsymbol
T_r(i)\bar{\boldsymbol g}(i)\boldsymbol x^H(i)}{\bar{\boldsymbol
x}^H(i)\bar{\boldsymbol g}(i)}]$;\\
$\tau(i)=P_e^2(i)[y^{\ast}(i)]^2\frac{[1-\frac{\delta(i)}{|y(i)|}]^2}{\bar{\boldsymbol
w}^H(i)\boldsymbol G_r^H(i)\boldsymbol x(i)\bar{\boldsymbol
x}^H(i)\bar{\boldsymbol g}(i)}$.

The last step is to give a $P_e(i)$ in order to provide good
estimates. In \cite{Lima}, a fixed probability that approximates the
update rate of the parameters is introduced. Here, we use a modified
version and involve the time-varying bound in our time-varying
probability, which is
\begin{equation}\label{35}
P_e(i)=2Q\big(\frac{\delta(i)}{\sigma_n}\big)+P_{\textrm{min}},
\end{equation}
where $Q(\cdot)$ is the complementary Gaussian cumulative
distribution given by
$Q(x)=\int_{x}^{\infty}\frac{1}{\sqrt{2\pi}}e^{-t^2/2}dt$, and
$P_{\textrm{min}}$ is prior information about the minimum update
rate for the proposed algorithm to reach a relatively high
performance. Both $P_e(i)$ and $P_{\textrm{min}}$ are different with
respect to different $N$ and reflect the data-selective updating
behavior as $N$ increases. The term $P_{\textrm{min}}$ is a certain
value if the number of snapshots is fixed.

The weight error vector can be calculated via its updating equation
\eqref{34}. The MSE value of the proposed JIO-SM-SG algorithm at a
given time instant $i$ can be obtained by substituting \eqref{34}
into \eqref{28}, which is
\begin{equation}\label{36}
\begin{split}
&J_{\textrm{mse}}(i+1)=J_{\textrm{min}}\\
&+\sigma_x^2\textrm{tr}\Big\{\textrm{cov}\big\{[\boldsymbol
I-\boldsymbol V(i)]\boldsymbol e_w(i)-\boldsymbol V(i)\boldsymbol
w_{\textrm{opt}}+\tau(i)\boldsymbol G_r(i)\bar{\boldsymbol
g}(i)\big\}\Big\}
\end{split}
\end{equation}
This analysis provides a means to predict the trend of the MSE
performance of the proposed algorithm. In the next section, we use
the simulation result to verify the validity of the analysis.

\subsection{Analysis of The Optimization Problem}
In this part, we provide an analysis based on the optimization
problem in (\ref{9}) and show a condition that allows the designer
to avoid local minima associated with the proposed optimization
problem. The analysis also considers the bounded constraint and the
constraint on the steering vector of the SOI to illustrate the
properties of the problem. Our analysis starts from the
transformation of the array output $y(i)$ in a more convenient form
and then substitutes its expression into the constrained
optimization problem to render the analysis. For simplicity, we drop
the time instant $i$ in the quantities.

From \eqref{1}, the array output can be written as:
\begin{equation}\label{37}
\begin{split}
y&=\bar{\boldsymbol w}^H\boldsymbol T_r^H\big\{\sum_{k=0}^{q-1}
\boldsymbol a(\theta_k)s_k+\boldsymbol n\big\}\\
&=\bar{\boldsymbol w}^H\sum_{k=0}^{q-1}\sum_{j=1}^{r}\boldsymbol
S_k\big[\boldsymbol t_j^H\boldsymbol
a(\theta_k)\boldsymbol\nu_j\big]+\bar{\boldsymbol w}^H\boldsymbol
I\sum_{j=1}^r\boldsymbol t_j^H\boldsymbol n\boldsymbol\nu_j,
\end{split}
\end{equation}
where $\boldsymbol S_k\in\mathbb C^{r\times r}$ is a diagonal matrix
with all its main diagonal entries equal to the transmitted data of
the $k$th user, i.e., $s_k$, $\boldsymbol t_j\in\mathbb
C^{m\times1}$ is the $j$th column vector of the transformation
matrix $\boldsymbol T_r$, and $\boldsymbol\nu_j\in\mathbb
C^{r\times1}$ is a vector containing a $1$ in the $j$th position and
zeros elsewhere. In order to proceed, we define {\small
\begin{equation}\label{38}
\begin{split}
\mathbb S_k=\begin{bmatrix} \boldsymbol 0 & \boldsymbol 0\\
\boldsymbol S_k & \boldsymbol 0\end{bmatrix}; \mathbb S_n=\begin{bmatrix} \boldsymbol 0 & \boldsymbol 0\\
\boldsymbol S_n & \boldsymbol 0\end{bmatrix}; \boldsymbol
f_k=\begin{bmatrix} \bar{\boldsymbol w}^{\ast}\\
\boldsymbol T_r^T\boldsymbol a_k^{\ast} \end{bmatrix};
\boldsymbol f_n=\begin{bmatrix} \bar{\boldsymbol w}^{\ast}\\
\boldsymbol T_r^T\boldsymbol n^{\ast} \end{bmatrix},
\end{split}
\end{equation}}
where $\mathbb S_k\in\mathbb C^{2r\times2r}$, $\mathbb S_n\in\mathbb
C^{2r\times2r}$ are the signal matrix containing the transmitted
signal of the $k$th user and the noise matrix containing
$\boldsymbol n$, respectively, and $\boldsymbol f_k\in\mathbb
C^{2r\times1}$ and $\boldsymbol f_n\in\mathbb C^{2r\times1}$ are the
parameter vector containing the steering vector of the $k$th user
and the noise vector related to $\boldsymbol n$, respectively.

According to \eqref{38}, the array output can be expressed by
\begin{equation}\label{39}
y=\sum_{k=0}^{q-1}\boldsymbol f_k^H\mathbb S_k\boldsymbol
f_k+\boldsymbol f_n^H\mathbb S_n\boldsymbol f_n,
\end{equation}
where we notice that, if $k=0$, it has $\boldsymbol f_0^H\mathbb
S_0\boldsymbol f_0=\bar{\boldsymbol w}^H\boldsymbol T_r^H\boldsymbol
a(\theta_0)s_0=\gamma s_0$, which is based on the constraint on the
steering vector of the SOI in \eqref{9}. The expression in
\eqref{39} involves this constraint in the array output and thus
will simplify the derivation when substituting into the optimization
problem. This is also the reason why we use \eqref{39} instead of
$y=\bar{\boldsymbol w}^H\boldsymbol T_r^H\boldsymbol x$ for the
analysis.

Before taking the analysis further, we use two assumptions. First,
the signals are assumed to be transmitted independently. Second, we
consider a noise free case \cite{Xu} or a sufficiently high
signal-to-noise ratio (SNR) condition. Under these assumptions,
using the Lagrange multiplier method, the optimization problem in
\eqref{9} can be written as
\begin{equation}\label{40}
\begin{split}
J(\boldsymbol T_r, \bar{\boldsymbol w})&=\mathbb
E[|y|^2]+2\lambda\mathfrak{R}[|y|^2-\delta^2]\\
&=\mathbb E[\sum_{k=0}^{q-1}\boldsymbol f_k^H\mathbb S_k\boldsymbol
f_k\boldsymbol f_k^H\mathbb S_k^H\boldsymbol
f_k]\\
&~~~+2\lambda\mathfrak{R}[\sum_{k=0}^{q-1}\sum_{l=0}^{q-1}\boldsymbol
f_k^H\mathbb S_k\boldsymbol f_k\boldsymbol f_l^H\mathbb
S_l^H\boldsymbol f_l-\delta^2],
\end{split}
\end{equation}
where the constraint on the steering vector of the SOI is not
included since it has been enclosed in \eqref{39}.

In order to evaluate the property of \eqref{40}, we can verify if
the Hessian matrix $\boldsymbol H_0$ \cite{Luenberger} with respect
to $\boldsymbol f_0$ of the desired user is positive semi-definite
for all nonzero vector $\boldsymbol u$ with $\boldsymbol
u^H\boldsymbol H_0\boldsymbol u\geq0$. Computing the Hessian of the
above optimization problem for the desired user we obtain
\begin{equation}\label{41}
\begin{split}
\boldsymbol H_0&=\frac{\partial}{\partial\boldsymbol
f_0^H}\frac{\partial J}{\partial\boldsymbol f_0}\\
&=\mathbb E\big[\boldsymbol H_{01}+\boldsymbol
H_{02}\big]+2\lambda\mathfrak{R}\big[\boldsymbol H_{01}+\boldsymbol
H_{02}+\boldsymbol H_{03}\big],
\end{split}
\end{equation}
where\\
$\boldsymbol H_{01}=\mathbb S_0\boldsymbol f_0\boldsymbol
f_0^H\mathbb S_0^H+\mathbb S_0^H\boldsymbol f_0\boldsymbol
f_0^H\mathbb S_0$;\\
$\boldsymbol H_{02}=\boldsymbol f_0^H\mathbb
S_0^H\boldsymbol f_0\mathbb S_0+\boldsymbol f_0^H\mathbb
S_0\boldsymbol f_0\mathbb
S_0^H$;\\
$\boldsymbol H_{03}=\sum_{k=1}^{q-1}\boldsymbol f_k^H\mathbb
S_k^H\boldsymbol f_k\mathbb S_k+\sum_{l=1}^{q-1}\boldsymbol
f_l^H\mathbb S_l\boldsymbol f_l\mathbb S_l^H$.
According to the constraint, $\boldsymbol
H_{02}=2\mathfrak{R}[\gamma^{\ast}s_0^{\ast}\mathbb S_0]$. From
\eqref{41}, $\boldsymbol H_{01}$ yields a positive semi-definite
matrix, while $\boldsymbol H_{03}$ is an undetermined term. Thus,
for avoiding the local minima associated with the optimization
problem, a sufficient condition is to guarantee
\begin{equation}\label{42}
\boldsymbol H_{0}'=\mathbb E[\boldsymbol
H_{02}]+2\lambda\mathfrak{R}[\boldsymbol H_{02}+\boldsymbol H_{05}],
\end{equation}
to be positive semi-definite, i.e., $\boldsymbol u^H\boldsymbol
H_0'\boldsymbol u\geq0$. This task can be achieved by selecting an
appropriate Lagrange multiplier $\lambda$. For the Hessian matrix
with respect to the other users, $\boldsymbol H_{k}$ ($k=1,\ldots,
q-1$), we could use the same way to avoid the local minima of the
optimization problem.

\section{Simulations}

In this section, we evaluate the performance of the proposed
JIO-SM-SG and JIO-SM-RLS adaptive algorithms for designing LCMV
beamformers and compare them with existing algorithms. Specifically,
we compare the proposed algorithms with the full-rank (FR) SG and
RLS algorithms \cite{Haykin} with/without the SMF technique, and the
reduced-rank algorithms based on the MSWF \cite{Honig2} and the AVF
\cite{Pados} techniques. In all simulations, we assume that there is
one desired user in the system and the related DOA is known
beforehand by the receiver. All the results are averaged by $1000$
runs. The input signal-to-interference ratio (SIR) is SIR=$-20$ dB.
We consider the binary phase shift keying (BPSK) modulation scheme
and set $\gamma=1$ for the studied algorithms. Simulations are
performed with a ULA containing $m=64$ sensor elements with
half-wavelength interelement spacing. We consider $m$ large in order
to show their advantages of the proposed algorithms in terms of
performance and computational complexity when the number of elements
in the beamformer is large.

\begin{figure}[htb]
\begin{minipage}[h]{1.0\linewidth}
  \centering
  \centerline{\epsfig{figure=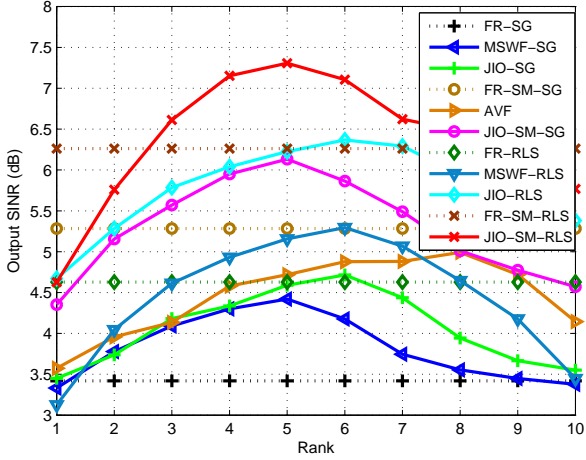,scale=0.6}} \vspace{-0.5em}\caption{Output SINR versus the
number of rank $r$.} \label{fig:rank}
\end{minipage}
\end{figure}

{  In Fig. \ref{fig:rank}, we assess the impact of the rank $r$ on
the output signal-to-interference-plus-noise ratio (SINR)
performance of the proposed and existing algorithms. There are
$q=25$ users in the system whose DoAs are generated with uniform
random variables between $0$ and $180$ degrees. The input SNR is
SNR=$10$ dB. In order to show the convergence behavior, we set the
total number of snapshots to $N=300$. The value of the rank $r$ is
chosen between $1$ and $10$ because reduced-rank adaptive algorithms
usually show the best performance for these values
\cite{Goldstein,Wang}. A reduced-rank algorithm with a lower rank
(e.g., $r=3$) often converges quickly whereas a reduced-rank
technique with a larger rank (e.g., $r=8$) reaches a higher SINR
level at steady state. We have chosen the parameters $\alpha=22$,
$\beta=0.99$, $\mu_T(1)=\mu_{\bar{w}}(1)=0.05$ for the proposed
JIO-SM-SG algorithm, and $\alpha=26$, $\beta=0.992$,
$\rho=1.3\times10^{-3}$, $\varrho=1.0\times10^{-4}$ for the proposed
JIO-SM-RLS algorithm in order to optimize the performance of the
algorithms. By changing these parameters, the performance will be
degraded, the update rate will decrease if the threshold $\delta$ is
large, whereas the update rate will increase if the threshold
$\delta$ is small. Note that $\lambda_1(i)$ should be in accordance
with the setting of the forgetting factor and thus
$0.1\leq\lambda_1(i)\leq0.998$ is used for implementation. Fig.
\ref{fig:rank} suggests that the most adequate rank for the proposed
algorithms to obtain the best performance in this example is $r=5$,
which is equal to or lower in comparison to the existing
reduced-rank algorithms. Besides, we also checked that this rank
value is rather insensitive to the number of users or interferers in
the system, to the number of sensor elements, and work efficiently
for the scenarios considered in the next examples. Since the best
$r$ is usually much smaller than the number of elements $m$, it
leads to a significant computational reduction. In general, we may
expect some variations for the optimal rank $r$ which should be in
the range $3<r<10$. In the following simulations, we use $r=5$ for
the proposed algorithms.}

\begin{figure}[htb]
\begin{minipage}[h]{1.0\linewidth}
  \centering
  \centerline{\epsfig{figure=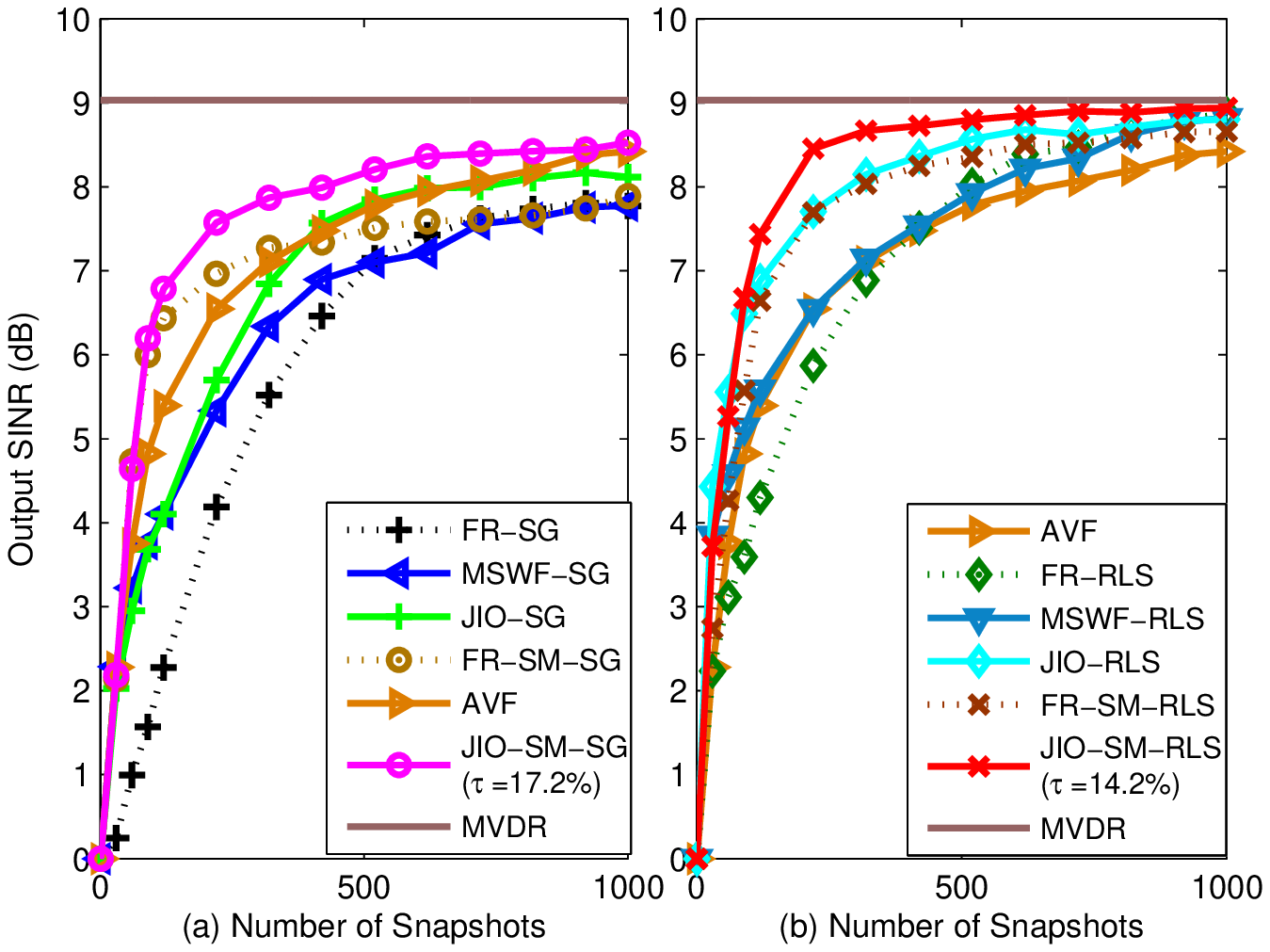,scale=0.6}} \vspace{-0.5em}\caption{Output SINR versus the
number of snapshots for (a) SG-based algorithms; (b) RLS-based
algorithms.} \label{fig:sinr}
\end{minipage}
\end{figure}

In Fig. \ref{fig:sinr}, we evaluate the SINR performance of the
proposed and existing algorithms versus the number of snapshots. It
includes two experiments, which compare the SG-based and the
RLS-based algorithms. The AVF algorithm is included in both
experiments to make a clear comparison. The scenario and the
coefficients for the proposed algorithms are the same as in Fig.
\ref{fig:rank}. The number of snapshots is $N=1000$. In Fig.
\ref{fig:sinr} (a), the JIO-based algorithms show a better
convergence rate than other full-rank and reduced-rank algorithms.
The proposed JIO-SM-SG algorithm has a good performance and only
requires $17.2\%$ updates ($172$ updates for $1000$ snapshots),
reducing the computational cost. Fig. \ref{fig:sinr} (b) exhibits a
similar result for the RLS-based algorithms. The JIO-SM-RLS
converges quickly to the steady-state, which is close to the minimum
variance distortionless response (MVDR) solution \cite{Trees}. The
update rate is $\tau=14.2\%$, which is much lower than its
reduced-rank counterparts that require $100\%$ updates.

\begin{figure}[htb]
\begin{minipage}[h]{1.0\linewidth}
  \centering
  \centerline{\epsfig{figure=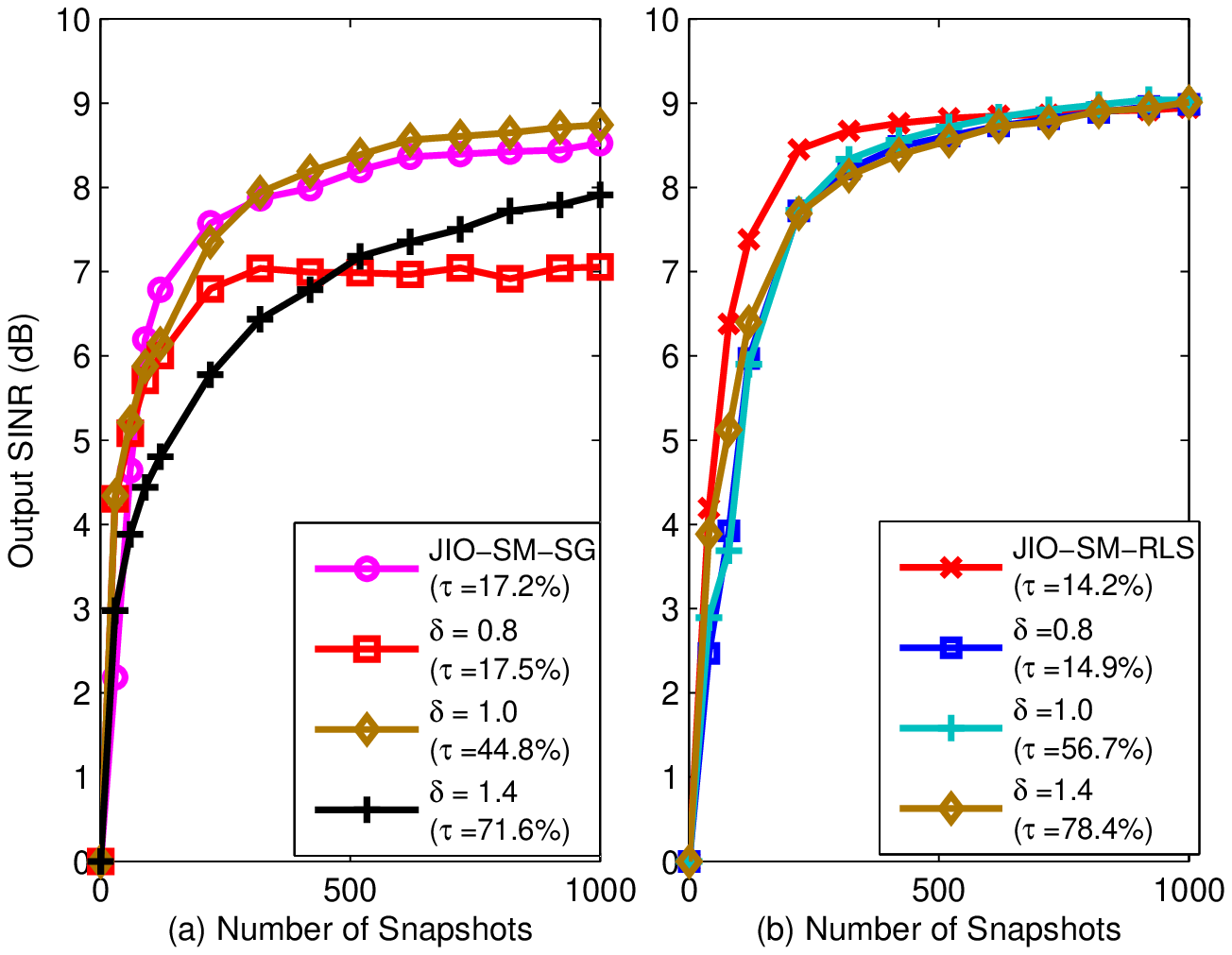,scale=0.6}} \vspace{-0.5em}\caption{Output SINR versus the
number of snapshots for (a) SG-based algorithms; (b) RLS-based
algorithms.} \label{fig:sm_fix}
\end{minipage}
\end{figure}

Fig. \ref{fig:sm_fix} shows the SINR performance of the proposed
algorithms with the fixed and time-varying bounds. The scenario is
the same as that in Fig. \ref{fig:rank}. From Fig. \ref{fig:sm_fix}
(a), we find that the curve with the fixed bound $\delta=1.0$ has
comparable SINR values to the proposed one as the number of
snapshots increases. The reason is that we use the BPSK modulation
scheme and thus the absolute value of the ideal array output should
equal $1$, which follows the constraint and achieves high SINR
values. However, it requires more updates ($\tau=44.8\%$) and has to
afford a much higher computational load. The curves with higher
($\delta=1.4$) or lower ($\delta=0.8$) bounds exhibit the worse
convergence performance. The proposed JIO-SM-SG algorithm with the
time-varying bound performs the data-selective updates to obtain a
good tradeoff between the complexity and the performance. The same
result can be found in Fig. \ref{fig:sm_fix} (b) for the proposed
JIO-SM-RLS algorithm, which uses even less updates to obtain an
enhanced performance.

%

\begin{figure}[htb]
\begin{minipage}[h]{1.0\linewidth}
  \centering
  \centerline{\epsfig{figure=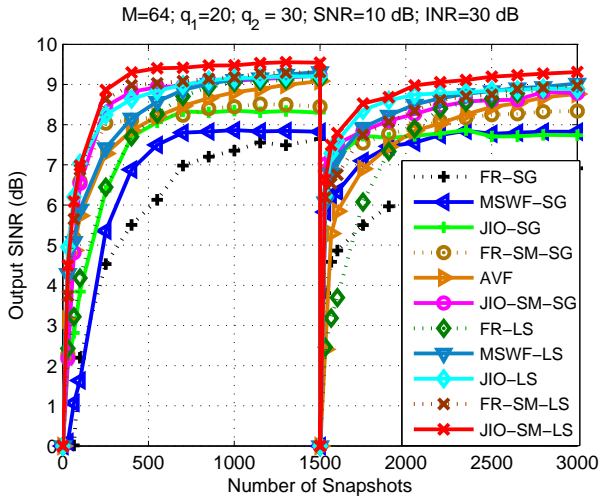,scale=0.72}} \vspace{-0.5em}\caption{Output SINR versus the
number of snapshots in dynamic scenario with additional users enter
and/or leave the system.} \label{fig:more}
\end{minipage}
\end{figure}

In the next experiment, we consider a non-stationary scenario,
namely, when the number of users changes in the system, and check
the tracking performance of the proposed algorithms. The system
starts with $q=20$ users including one desired user. The
coefficients are $r=5$, $\alpha=18$, $\beta=0.99$,
$\mu_{T}(1)=\mu_{\bar{w}}(1)=0.05$ for the proposed JIO-SM-SG
algorithm and $r=5$, $\alpha=19$, $\beta=0.995$,
$\rho=1.3\times10^{-3}$, $\varrho=1\times10^{-4}$ for the proposed
JIO-SM-RLS algorithm. From Fig. \ref{fig:more}, the proposed
algorithms achieve a superior convergence performance to the other
compared algorithms. The environment experiences a sudden change at
$i=1500$. We have $10$ interferers entering the system. This change
degrades the SINR performance for all the algorithms. The proposed
algorithms track this change and converge rapidly to the
steady-state since the data-selective updates reduce the number of
parameter estimation and thus keep a faster convergence rate.
Besides, the time-varying bound provides information for them to
follow the changes of the scenario. It is clear that the proposed
algorithms still keep low update rates ($\tau=15.3\%$ for the
JIO-SM-SG and $\tau=16.0\%$ for the JIO-SM-RLS even under
non-stationary conditions.

\begin{figure}[htb]
\begin{minipage}[h]{1.0\linewidth}
  \centering
  \centerline{\epsfig{figure=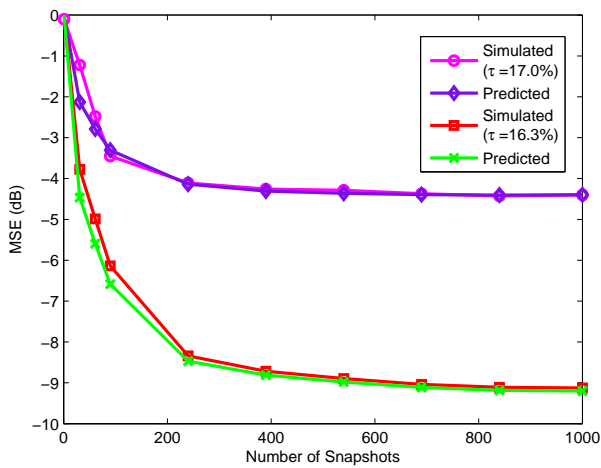,scale=0.6}} \vspace{-0.5em}\caption{MSE performance versus the
  number of snapshots for the proposed JIO-SM-SG algorithm.} \label{fig:mse}
\end{minipage}
\end{figure}

{  In the last experiment, the simulated and analytical results in
Subsection V-C for the proposed JIO-SM-SG algorithm are compared.
The simulated curves are obtained via simulations in Fig.
\ref{fig:mse} and the predicted ones are from \eqref{36}}. In this
scenario, there are $q=20$ users in the system and INR$=25$ dB. We
compare the results with two different SNR values, i.e., SNR$=5$ dB
and SNR$=10$ dB. The coefficients are $\alpha=9.7$, $\beta=0.99$,
$\mu_{T}(1)=\mu_{\bar{w}}(1)=0.05$, and $r=5$. To get the predicted
MSE, we set $P_{\textrm{min}}=17.0\%$ (for SNR=$5$ dB) and
$P_{\textrm{min}}=16.3\%$ (for SNR=$10$ dB), which are in accordance
with the update rates of the simulated MSE and provide a fair
comparison. From Fig. \ref{fig:mse}, the predicted curves agree with
the simulated ones, especially when $i\geq200$, verifying the
validity of our analysis. Note that there is a small gap between the
simulated and predicted curves at the beginning. The reason is that
the number of snapshots is insufficient for the proposed JIO-SM-SG
algorithm to provide accurate estimates if $i\leq m$ (or before the
algorithm converges to the steady-state). The update rates for the
simulated curves are quite low and thus decrease the computational
cost for the proposed algorithm.

\section{Concluding Remarks}

We have introduced a new reduced-rank framework that incorporates
the SMF technique into the reduced-rank JIO scheme for beamforming.
According to this framework, we have considered reduced-rank LCMV
designs with a bounded constraint on the amplitude of the array
output, and developed SG-based and RLS-based adaptive algorithms for
beamforming. The proposed algorithms have employed the received data
to construct a space of feasible solutions for the updates. They
have a superior convergence and an enhanced tracking performance
over their existing counterparts due to the iterative exchange of
information between the transformation matrix and the reduced-rank
weight vector. In addition, the proposed algorithms can save
computational costs due to the data-selective updates. A
time-varying bound was employed to adjust the step size values for
the SG-based algorithm and the forgetting factor values for the
RLS-based algorithm, making the proposed algorithms more robust to
dynamic scenarios. The results have shown the advantages of the
proposed algorithms and verified the analytical formulas derived.

\begin{appendix}
{
\section*{Derivation of JIO-SG Algorithm}
The recursions for the JIO-SG algorithms are derived from
(\ref{10}). Fixing $\bar{\boldsymbol w}$ and computing the gradient
terms of (\ref{10}) with respect to ${\boldsymbol T}_r$ and using a
gradient descent rule \cite{Diniz}, we have
\begin{equation}
{\boldsymbol T}_r[i+1] = {\boldsymbol T}_r[i]- \mu_T ( y^*[i]
{\boldsymbol x} [i] \bar{\boldsymbol w}^H[i] + 2 \eta {\boldsymbol
a}(\theta_0) \bar{\boldsymbol w}^H[i]). \label{sg_gd}
\end{equation}
Substituting the above into the constraint $\bar{\boldsymbol
w}^H[i]{\boldsymbol T}_r^H[i] {\boldsymbol a}(\theta_0) = \gamma$,
we obtain the value of the Lagrange multiplier
\begin{equation}
\eta = 1/2 ({\boldsymbol a}^H(\theta_0){\boldsymbol a}
(\theta_0))^{-1} {\boldsymbol a}^H(\theta_0) {\boldsymbol x}[i]
y^*[i]
\end{equation}
Substituting $\eta$ into (\ref{sg_gd}), we obtain (\ref{11}). The
recursion for $\bar{\boldsymbol w}$ is obtained by an analogous
gradient descent rule
\begin{equation}
\bar{\boldsymbol w}[i+1] = \bar{\boldsymbol w}[i]- \mu_w ( y^*[i]
{\boldsymbol T}_r^H[i]{\boldsymbol x} [i]  + 2 \eta {\boldsymbol
T}_r^H[i]{\boldsymbol a}(\theta_0) ). \label{sg_gd2}
\end{equation}
Using the constraint again with the above recursion, we can obtain
the value for the Lagrange multiplier for use in the update of
$\bar{\boldsymbol w}$ and which results in (\ref{13}).}

\section*{Derivation of Variable Step Size Values}
In this appendix, we derive the expressions in (\ref{13}) and
(\ref{14}). We drop the time instant $i$ for simplicity. According
to the optimization problem, substituting (\ref{11}) into the
bounded constraint in (\ref{9}), we have
\begin{equation}\label{b_1}
\Big|\bar{\boldsymbol w}^H\Big\{\boldsymbol
T_r^H-\mu_{T}^{\ast}y\big[\bar{\boldsymbol w}{\boldsymbol
x}^H-{\bar{\boldsymbol w}\boldsymbol x^H\boldsymbol
a(\theta_0)\boldsymbol a^H(\theta_0)}\big]\Big\}\boldsymbol
x\Big|=\delta
\end{equation}

The above equation can be expressed in an alternative form, which is
\begin{equation}\label{b_2}
\Big|y-\mu_T^{\ast}y\bar{\boldsymbol w}^H\bar{\boldsymbol
w}\boldsymbol x^H\big[\boldsymbol I-\boldsymbol
a(\theta_0)\boldsymbol a^H(\theta_0)\big]\boldsymbol x\Big|=\delta
\end{equation}

Making an arrangement to \eqref{b_2}, we obtain the variable step
size expression for the transformation matrix in \eqref{13}.

Also, substituting (\ref{12}) into the bounded constraint, we obtain
\begin{equation}\label{b_3}
\Big|\bar{\boldsymbol w}^H\bar{\boldsymbol
x}-\mu_{\bar{w}}^{\ast}y\bar{\boldsymbol x}^H[\boldsymbol
I-\frac{\bar{\boldsymbol a}(\theta_0)\bar{\boldsymbol
a}^H(\theta_0)}{\bar{\boldsymbol a}^H(\theta_0)\bar{\boldsymbol
a}(\theta_0)}]\bar{\boldsymbol x}\Big|=\delta,
\end{equation}
where the expression of $\mu_{\bar{w}}$ in (\ref{14}) can be
obtained by performing mathematical transformations to \eqref{b_3}.

\section*{Derivation of (\ref{16})}
In this appendix, we derive the expression of the transformation
matrix in (\ref{16}). Given $\bar{\boldsymbol w}(i)\neq\boldsymbol
0$, taking the gradient of (\ref{15}) with respect to $\boldsymbol
T_r(i)$, we have
\begin{equation}\label{a_1}
\begin{split}
\nabla J_{\boldsymbol
T_r(i)}=&\sum_{l=1}^{i-1}\lambda_1^{i-l}(i)\boldsymbol
x(l)\boldsymbol x^H(l)\boldsymbol T_r(i)\bar{\boldsymbol
w}(i)\bar{\boldsymbol w}^H(i)\\
&+\lambda_1(i)\boldsymbol x(i)\boldsymbol x^H(i)\boldsymbol
T_r(i)\bar{\boldsymbol w}(i)\bar{\boldsymbol
w}^H(i)+\lambda_2\boldsymbol a(\theta_0)\bar{\boldsymbol w}^H(i).
\end{split}
\end{equation}

Making $\nabla J_{\boldsymbol T_r(i)}=\boldsymbol 0$ and
right-multiplying the both sides by $\bar{\boldsymbol w}(i)$, and
rearranging the expression, it becomes {\small
\begin{equation}\label{a_2}
\boldsymbol T_r(i)\bar{\boldsymbol
w}(i)=-\lambda_2\big[\sum_{l=1}^{i-1}\lambda^{i-l}_1(i)\boldsymbol
x(l)\boldsymbol x^H(l)+\lambda_1(i)\boldsymbol x(i)\boldsymbol
x^H(i)\big]^{-1}\boldsymbol a(\theta_0).
\end{equation}}

Considering the assumption $\lambda_1(i)\rightarrow1$ and using the
matrix inversion lemma, we have
\begin{equation}\label{a_3}
\begin{split}
\boldsymbol T_r(i)\bar{\boldsymbol w}(i)&=-\lambda_2\big[\boldsymbol
R(i-1)+\lambda_1(i)\boldsymbol x(i)\boldsymbol
x^H(i)\big]^{-1}\boldsymbol a(\theta_0)\\
&=-\lambda_2\boldsymbol P(i)\boldsymbol a(\theta_0),
\end{split}
\end{equation}
where $\boldsymbol P(i)$ has been given in (\ref{18}).

Let $\boldsymbol v(i)=\boldsymbol P(i)\boldsymbol a(\theta_0)$, the
solution of $\boldsymbol T_r(i)$ can be regarded to find the
solution to the linear equation
\begin{equation}\label{a_4}
\boldsymbol T_r(i)\bar{\boldsymbol w}(i)=\boldsymbol v(i),
\end{equation}
where there exists multiple $\boldsymbol T_r(i)$ satisfying this
equation if only $\bar{\boldsymbol w}\neq\boldsymbol 0$. We derive
the minimum Frobenius-norm solution for stability. We write
$\boldsymbol T_r(i)$ and $\boldsymbol v(i)$ in the form of
\begin{equation}\label{a_5}
\boldsymbol T_{r}(i)=\begin{bmatrix}
 \bar{\boldsymbol t}_{1}(i)\\
 \bar{\boldsymbol t}_{2}(i)\\
 \vdots\\
 \bar{\boldsymbol t}_{m}(i)\\ \end{bmatrix};~~
{\boldsymbol v}(i)=\begin{bmatrix}
 v_{1}(i)\\
 v_{2}(i)\\
 \vdots\\
 v_{m}(i)\\ \end{bmatrix},
\end{equation}
where $\bar{\boldsymbol t}_j\in\mathbb C^{r\times1}$ with $j=1,
\ldots, m$ denotes the row vector of the transformation matrix.
Thus, the search of the minimum Frobenius-norm solution is
simplified to the following $m$ subproblems:
\begin{equation}\label{a_6}
\textrm{minimize}~\|\bar{\boldsymbol
t}_j(i)\|^2,~~\textrm{subject~to}~~\bar{\boldsymbol
t}_j(i)\bar{\boldsymbol w}(i)=v_j(i).
\end{equation}

Solving the constrained optimization problem in (\ref{a_6}), we have
\begin{equation}\label{a_7}
\bar{\boldsymbol t}_j(i)=v_j(i)\frac{\bar{\boldsymbol
w}^H(i)}{\|\bar{\boldsymbol w}(i)\|^2}.
\end{equation}

Substituting (\ref{a_7}) into (\ref{a_4}) and considering the
definition of $\boldsymbol v(i)$, the minimum Frobenius-norm
solution is given by
\begin{equation}\label{a_8}
\boldsymbol T_r(i)=-\lambda_2\boldsymbol P(i)\boldsymbol
a(\theta_0)\frac{\bar{\boldsymbol w}^H(i)}{\|\bar{\boldsymbol
w}(i)\|^2},
\end{equation}
where $\lambda_2$ can be obtained by incorporating (\ref{a_3}) into
the constraint with respect to $\boldsymbol a(\theta_0)$, which is
\begin{equation}\label{a_9}
\lambda_2=-\frac{\gamma}{\boldsymbol a^H(\theta_0)\boldsymbol
P(i)\boldsymbol a(\theta_0)}.
\end{equation}
\end{appendix}

\end{document}